\documentclass{ws-p9-75x6-50}

\begin{document}

\title{Highlights of RXTE Studies of Compact Objects after $\sim$5 years}

\author{Hale Bradt}

\address{Massachusetts Institute of Technology, Room 37-587, Cambridge MA 02139-4307 USA\\E-mail: bradt@mit.edu}

\author{Richard E. Rothschild}

\address{University of California, San Diego, CASS, La Jolla, CA 92093-0424 USA\\E-mail: rrothschild@ucsd.edu}  

\author{Jean H. Swank}

\address{Goddard Space Flight Center, NASA, Code 666, Greenbelt MD 20771 USA\\E-mail: swank@pcasun1.gsfc.nasa.gov}  


\maketitle

\abstracts{
Observations with the Rossi X-ray Timing Explorer (RXTE) have led to fundamental 
progress in the study of compact objects, in particular neutron stars and black holes. In this 
paper we present briefly some highlights from the first $\sim$5 years of RXTE operations. }

\section {Introduction}
The Rossi X-ray Timing Explorer (RXTE), launched 1995 Dec. 30, is still operating and is expected to remain active for several more years. The mission is designed to probe the nature of compact objects through timing of the variable intensities and the tracking of broad-band spectra~\cite{br93}. The features that provide this capability are: 

\indent{(1) a large aperture Proportional Counter Array (PCA) which provides high statistics 
in short periods. The brighter sources can be studied with millisecond time resolutions 
comparable to the orbital time scales of matter close to neutron stars or stellar black holes}; 

\indent{(2) the capability to detect transient phenomena with the All Sky Monitor (ASM) 
together with spacecraft systems that permit rapid pointing response, within a few hours;}

\indent{(3) a broad spectral response with the PCA (2 $-$ 60 keV) together with the 
coaligned High Energy X-ray Timing Experiment (HEXTE; 15 $-$ 250 keV).}

\begin{figure}[thb]
\psfig{figure=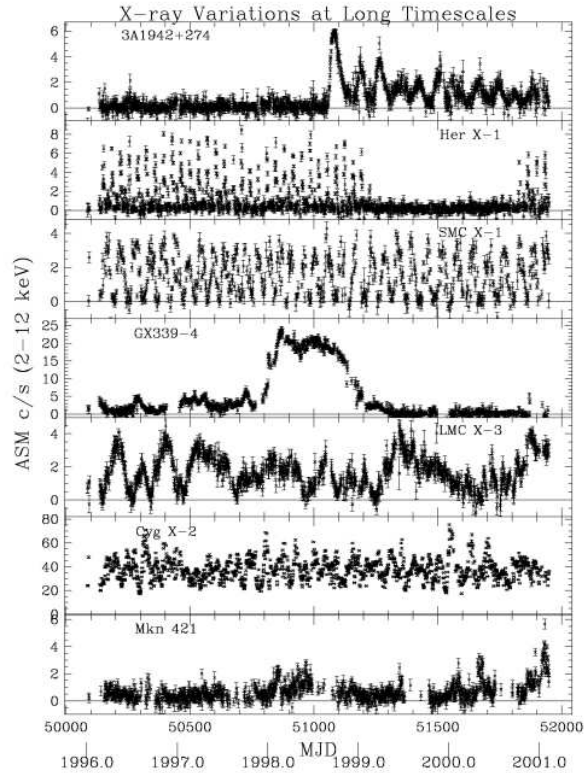,height=5.0in} 
\caption{ASM light curves of 7 quasi-persistent sources. The top three are neutron star systems, showing (from the top) a long orbital period, and superorbital periods of 35 and 60 d 
(respectively) that are most likely accreting precession disks. The other curves are (in 
descending order) a black hole candidate that entered a sustained soft spectral state, a black hole system with variability on time scales $>100$ d, a neutron star which shows $\sim$70--d quasiperiodic variability, and finally a flaring extragalactic blazar, from Remillard (pvt. comm.).
\label{asmpr}}
\end{figure}

\begin{figure}[thb]
\psfig{figure=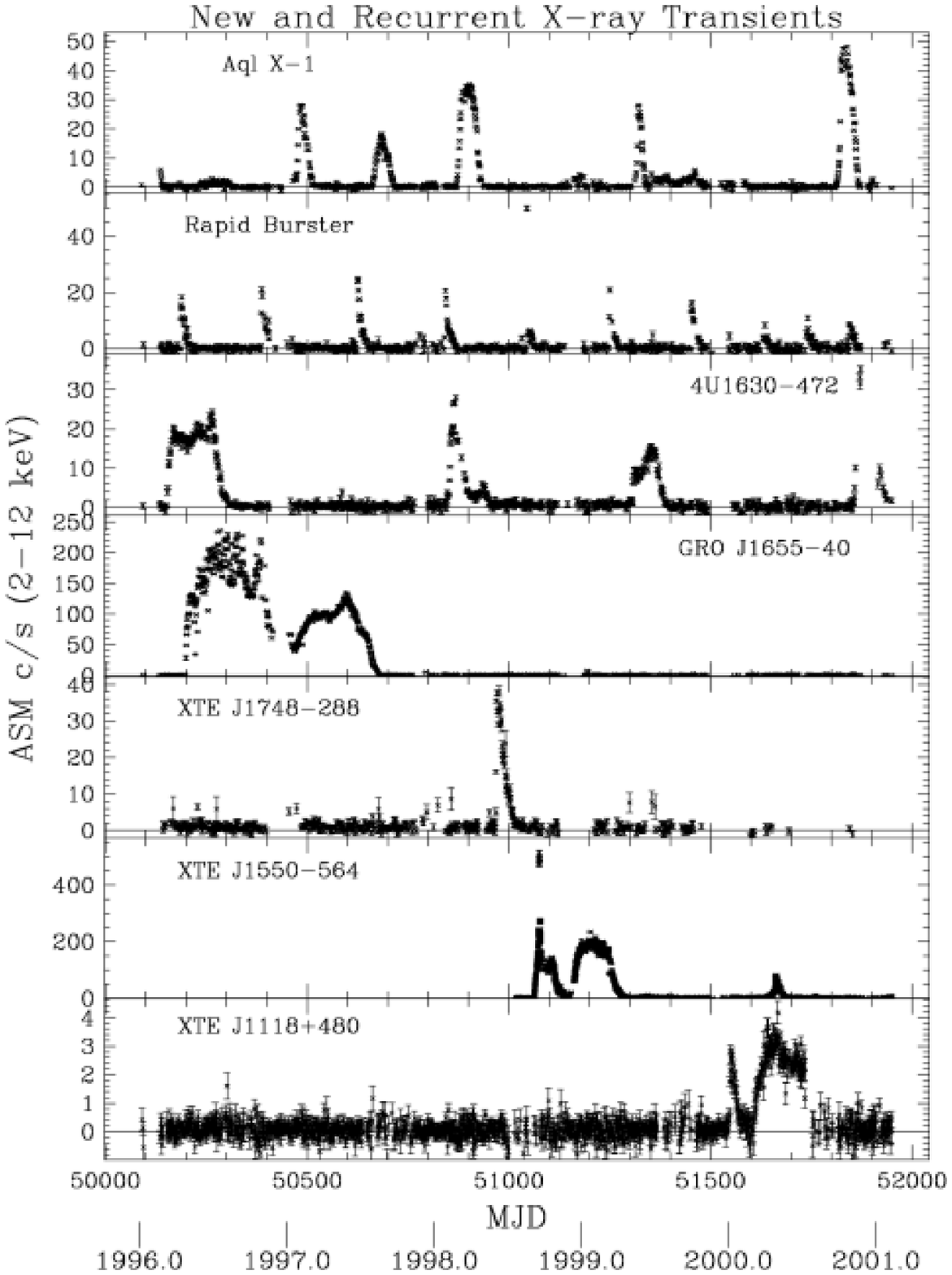,height=5.0in} 
\caption{ASM 5-year light curves (1.5 --12 keV) of (from top) two neutron star and 5 black-
hole systems, from Remillard (pvt . comm.). 
\label{asmtr}}
\end{figure}

The richness of variability in the x-ray sky is clear from the 5-year light curves of fourteen 
sources obtained from the ASM. Seven quasi-persistent sources are shown in Fig.~\ref{asmpr} 
and seven transients in Fig.~\ref{asmtr}.

In the spring of 2000, members of the RXTE Users Group, our
colleagues on the instrument teams, and other RXTE users, helped in
making the case for continued RXTE observations. The achievements of
the past 5 years of operations were important to that case. In this report, we outline highlights pertaining to compact objects, primarily neutron stars and black holes. We also touch upon progress in cataclysmic variables and the role of RXTE in the study of gamma ray bursts. 

This brief snapshot with sample references is neither complete nor necessarily balanced. It 
points up the broad themes of progress in compact-object physics, but it does not give
a proper view of the depth and richness in detail that RXTE is bringing to these topics. The 
reader is referred to reviews on individual topics, e.g., van der Klis~\cite{vdK00} on kilohertz 
oscillations in neutron star systems, Remillard~\cite{remBH} on black hole systems, Bradt 
et al.~\cite{br00} on transients, Nandra~\cite{na01} on active galactic nuclei, Mirabel \& Rodriguez~\cite{mir99} on microquasars, van Paradijs, Kouveliotou \& Wijers~\cite{vaP00} on gamma-ray burst afterglows, Mereghetti~\cite{mere01} and Thompson~\cite{th00} on anomalous x-ray pulsars and soft gamma-ray repeaters.

\section{Measurements in Regimes of Strong Gravity}
RXTE has obtained a diverse set of quantitative results that provide information on the 
regime of strong gravity found near neutron stars and stellar black holes. The signatures are 
both temporal and spectral. They arise when x-ray emitting gases approach the innermost 
stable circular orbit (ISCO) predicted in General Relativity (GR).

Rapid quasiperiodic oscillations (QPO) in x-ray flux (hundreds of Hz to 1.3 kHz) are found in both 
neutron star and black-hole systems. They must arise from the innermost regions of 
accretion. Candidate processes for these QPO include Kepler orbits, disk oscillations~\cite{no97}, periastron precession~\cite{st99}, frame dragging (nodal precession)~\cite{cu98}~\cite{mer99}, and beat frequencies~\cite{mi98}~\cite{la01}. 

\begin{figure}[phb]
\psfig{figure=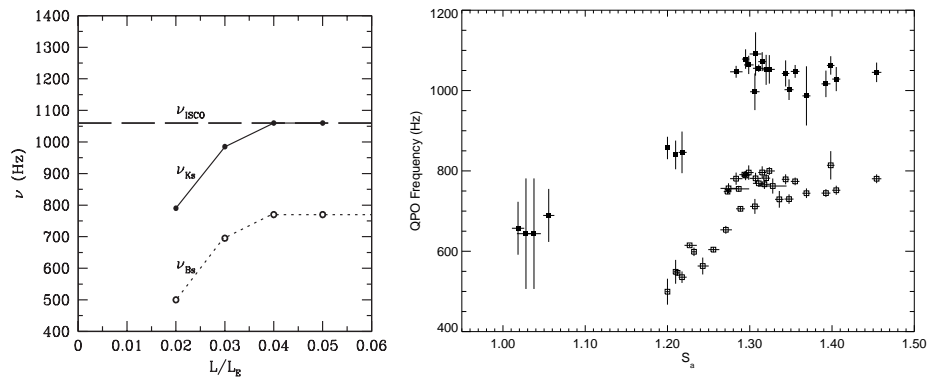,height=2.0in} 
\caption{Left: Prediction of behavior of QPO frequency, from Miller et al. (ref. 43). The two frequencies (upper and lower curves) are expected to saturate as accretion increases due to the presence of the innermost stable circular orbit (ISCO). Right: QPO frequency of 1820$-$30 versus position in color-color diagram, from Bloser et al. (ref. 4).
\label{1820}}
\end{figure}

\begin{figure}[phb]
\psfig{figure=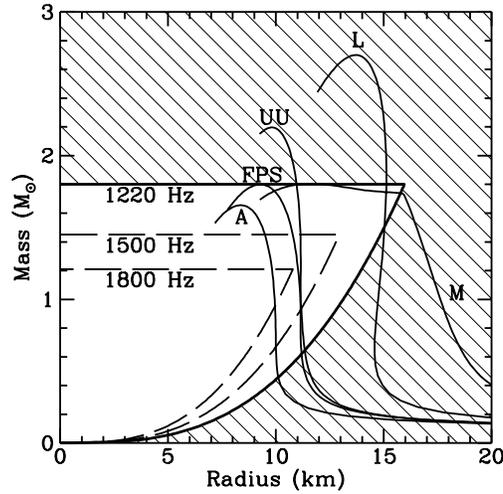,height=2.5in} 
\caption{Region of radius-mass plane for non-rotating neutron allowed by observation of 
1220-Hz QPO, with various equations of state, from Miller et al. (ref. 43).
\label{eqs}}
\end{figure}

Evidence for the ISCO is found in the temporal variability of the x-ray flux of the neutron star binary system 4U~1820$-$30. The observed behavior of the kHz oscillations~\cite{zh98}~\cite{blo00} as a function of position on the color-color diagram (presumably a measure of mass accretion rate) matches well the predicted behavior~\cite{mi98} (Fig.~\ref{1820}), though other factors could mimic this result~\cite{vdK00}. Kilohertz oscillations in neutron star systems studied with RXTE (see below) also constrain the equation of state of the neutron 
star~\cite{mi98}; Fig.~\ref{eqs}.

\begin{figure}[thb]
\psfig{figure=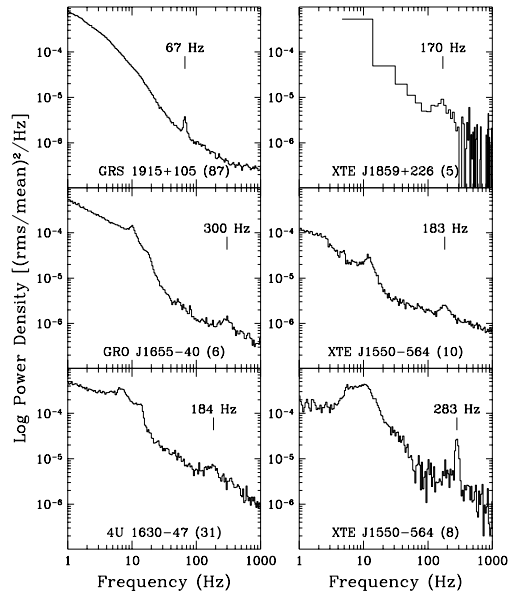,height=3.0in} 
\caption{High frequency QPO's in 5 black hole candidates. The number of summed PCA 
observations is given in parentheses, from Remillard (ref. 59). 
\label{bhosc}}
\end{figure}

High frequency oscillations found in five stellar black holes are shown in Fig.~\ref{bhosc}. A 
given model of the oscillations, in conjunction with the determination of black-hole mass via optical techniques, can yield the angular momentum parameter of the black hole. For example, the 300 Hz oscillations in GRO J1655$-$40 require maximal rotation if due to frame dragging~\cite{cu98} and negligible rotation if they are the Kepler frequency of the ISCO~\cite{rem300}.

RXTE spectra can tell us how closely accretion disks approach the compact object, the closer the hotter. The radius of the ISCO depends upon the spin parameter and mass of the compact object in GR. Thus, if the spectrally determined radius is, in fact, the radius of the ISCO, it provides another measure of the angular momentum parameter if the mass is independently known from dynamical measures. In the case of microquasars, the black hole may have to be nearly maximally rotating to be consistent with spectral results~\cite{zh97}. 

\section{Black hole systems}
\subsection{Spectra and low-frequency oscillations in black-hole systems}
X-ray spectra of black holes typically exhibit a thermal component usually considered to be 
from the accretion disk, and also a power-law component. There are two types of power law associated with 
different intensity states of the black hole~\cite{gro98}. Their natures are a mystery, though 
generally they are considered to arise from inverse-Compton upscattering from a hot plasma of electrons. In the ``low-hard state'' of black hole systems, the power law could arise from hot matter undergoing infall into the black hole, i.e., advection dominated flow~\cite{na95}, or it could arise from compact magnetic flares on the surface of the accretion disk~\cite{po99}. 

The relative behavior of the power law, the disk component, and the (sometimes extremely 
strong) low-frequency quasi periodic oscillations (QPOs) has been studied with RXTE 
through huge state variations in a number of black-hole systems. Patterns of behavior are 
emerging. For example, in GRS 1915+105, the low frequency QPOs are associated primarily 
with increases in the strength of the power law component~\cite{mun99}. Thus the 
oscillations are not solely a disk phenomenon. Current modeling efforts incorporate both the 
temporal and spectral behaviors in these systems~\cite{kaz99}~\cite{hua99}~\cite{now99}~\cite{po99}.

\subsection{Formation of relativistic astrophysical jets}
Microquasars, so called because of their superluminal radio 
emissions~\cite{mir94}~\cite{fe99}, are also highly variable x ray sources. As such, they 
provide important probes of the disk/jet connection. 

\begin{figure}[thb]
\psfig{figure=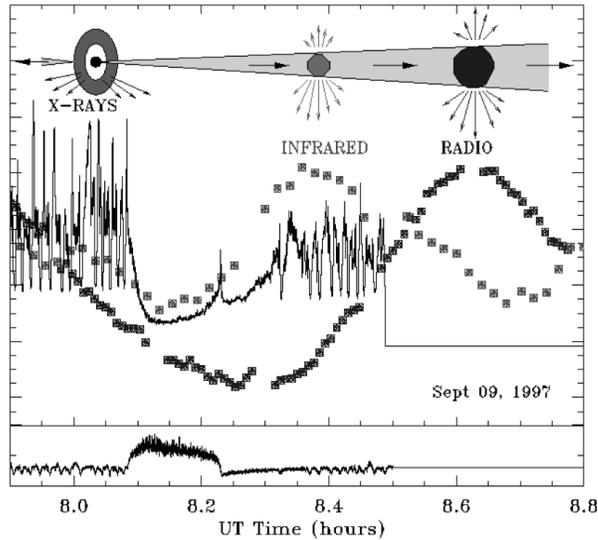,height=4.0in} 
\caption{Associated radio, infrared and x-ray light outbursts from microquasar GRS 
$1915+105$. The pattern repeats about every 30 minutes for hours or a few days. The lower 
plot is the x-ray hardness ratio, 13--30 keV / 2--13 keV, from Mirabel et al. (ref. 45).
\label{1915}}
\end{figure}

The microquasar GRS J1915+105 exhibits $\sim$10 dramatic types of x-ray modulation or 
oscillations on time scales of minutes~\cite{mun99}. Each of these modes may continue for a 
few hours or days. One of these modes exhibits a $\sim$30 minute oscillation period, and 
each oscillation has been found to be nearly coincident with a non-thermal radio/infrared 
outburst~\cite{poo97}~\cite{ei98}~\cite{mir98}; Fig.~\ref{1915}. This reveals directly the 
behavior of the accretion disk as jet outbursts occur. Changes in the x-ray spectrum imply the destruction of the inner, x-ray emitting part of an accretion disk~\cite{be97}. This disk material may become the ejecta in the recurring outbursts. These radio events have much smaller 
fluxes (0.01 -- 0.1 Jy) than the large events (0.5 -- 2.0 Jy) that have been mapped with VLBI 
to reveal superluminal motion. RXTE has yet to capture one of these rare larger events. 

The x-ray spectral state of the black hole system GX 339$-$4 (Fig.~\ref{asmpr}) is strongly correlated with the 
presence or absence of steady radio emission~\cite{fe99a}~\cite{cor00}. The steady radio 
emission may represent a steady jet. Extended emission from Cyg X$-$1 may also be such a 
jet~\cite{sti97}~\cite{bro99}. A steady jet in GRS 1915+105 has been imaged with VLBI on 
dimensions from 10 to 500 AU~\cite{dh00}. These investigations are attempting to relate the 
power-law of the low-hard state (index $\sim$1.7) to Compton scattering at the 
base of a steady jet.

\subsection{Spectra and variability in AGN}
RXTE is uniquely capable of producing well-sampled long-term x-ray light curves for multi-
wavelength monitoring campaigns of AGN. Monitoring campaigns coordinated with optical 
and ultraviolet observations have found patterns that imply a close coupling between the origins of the different bands, despite a lack 
of simple correlation in some cases~\cite{na01}.

RXTE has confirmed that radio galaxies have weaker reflection components and Fe lines than do radio quiet Seyfert galaxies~\cite{rot99}~\cite{era00}. This suggests that the presence of the jet may alter the inner environment of AGN. 

\begin{figure}[thb]
\psfig{figure=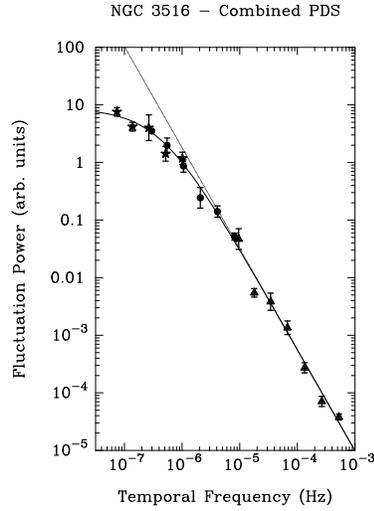,height=2.5in} 
\caption{Broadband power density spectrum of the Seyfert 1 AGN, NGC 3516. The cutoff corresponds to a period of 
$\sim$1 month, from Edelson \& Nandra (ref. 19).
\label{3516}}
\end{figure}

Monitoring campaigns covering several years for the first time
clearly demonstrate that the typical variability time scale for
Seyfert 1 galaxies ranges from weeks to months, as indicated, for example, by the turnover of the PDS of NGC 3516; Fig.~\ref{3516}~\cite{ed99}. Similar turnovers in stellar black-hole systems are found at much smaller time scales, e.g., $\sim$10 s for Cyg X$-$1. The ratio of the time 
constants of AGN to stellar black holes may reflect the relative black hole masses, e.g. 
$\sim$10 $M_{sun}$ for Cyg X-1 and $10^{6} - 10^{7} M_{sun}$ for NGC 3516~\cite{ed99}. The origin of these time scales is not known. 

\begin{figure}[thb]
\psfig{figure=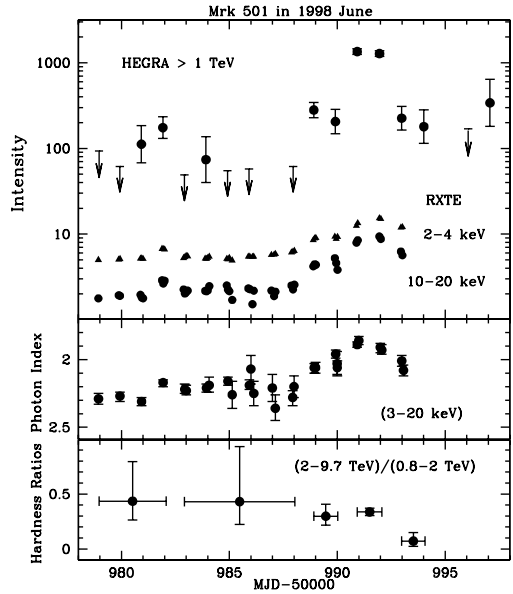,height=3.0in} 
\caption{TeV and x-ray associated flares from blazar Mk 501, from Sambruna et al. (ref. 62).
\label{tev}}
\end{figure}

Observations of the brighter BL Lac objects show substantial variability in ASM data; see 
Mk 421 in Fig.~\ref{asmpr}. Coordinated TeV and RXTE/PCA observations of Mk 501 
have shown dramatically clear simultaneous flares~\cite{sa00}; Fig.~\ref{tev}. The 
data are consistent with the synchrotron self Compton model wherein the 
electrons that give rise to the (x-ray) synchrotron photons upscatter these same photons to 
TeV energies. 

Iron lines and Compton reflection components have been measured in
spectra of individual Seyfert 1 and Seyfert 2 galaxies. In one case, MCG -6-30-15, the iron 
line does not share the variations of the continuum~\cite{rey00}. This challenges the 
interpretation of broadened lines originating in the inner disk.

\section{Neutron star systems}
\subsection{Spin-up of neutron stars}
The RXTE discovery of kilohertz oscillations in low-mass x-ray binary systems has provided 
a powerful new tool for the exploration of the regions very close to the neutron star. The 
frequencies, ranging up to about 1300 Hz clearly are comparable to those expected from 
Kepler motion in the ISCOs or from very rapid spinning of the neutron star. The frequency patterns and variations exhibit clear consistent patterns that provide strong 
diagnostics. See the above mentioned review~\cite{vdK00}.

\begin{figure}[thb]
\psfig{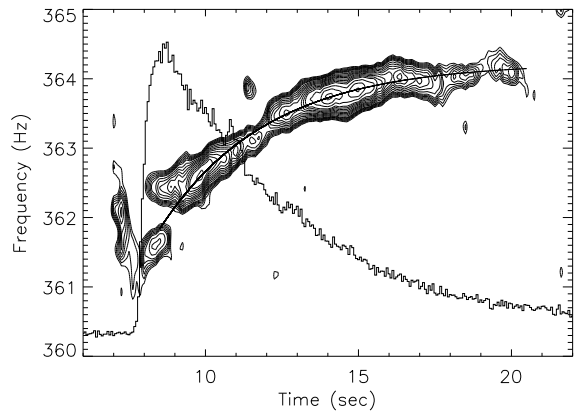} 
\caption{X-ray burst in LMXB 1728$-$34 (histogram) with frequency of QPO (contours), 
from Strohmayer (private communication); see ref. 68.
\label{bstosc}}
\end{figure}

The oscillations indicate that the neutron stars in low-mass x-ray binaries (LMXB) 
are rotating between 300 and 600~Hz, as is expected if they have been accreting for $10^8$--
$10^9$~yr in binaries with low-mass companions. Quasi-coherent oscillations during x-ray 
bursts (a runaway thermonuclear flash on the neutron star surface) are most likely a direct 
indicator of the neutron star spin when the frequency settles down to its coherent value; Fig.~\ref{bstosc}~\cite{str96}.

\begin{figure}[thb]
\psfig{figure=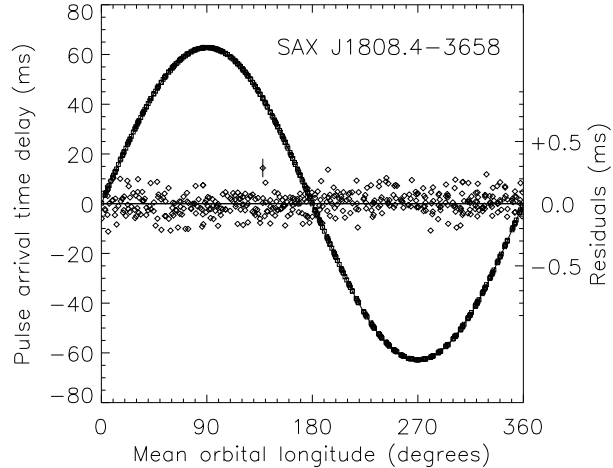,height=2.5in} 
\caption{Doppler orbit from PCA data of the only known accreting millisecond pulsar, 
SAX~J1808.4--3658, from Chakrabarty \& Morgan (ref. 9).
\label{1808}}
\end{figure}

One, and so far only one, transient LMXB exhibits truly coherent pulsations, SAX~J1808.4--
3658 at 401~Hz~\cite{wi98}~\cite{ch98}. This 2.5-ms oscillation surely is the spin 
period of the neutron star. Doppler shifts are tracked to obtain with high precision the 2.01-h orbit; Fig.~\ref{1808}. This is the first discovered {\it accreting} millisecond pulsar. It 
establishes the fact that millisecond radio pulsars are spun up due to the accretion torques in x-ray binaries. The magnetic fields of these old accreting neutron stars are in the range $10^8$--$10^9$~G, the same as those of recycled radio pulsars.

Accumulating evidence from studies of the high frequency oscillations
points to old accreting neutron stars having 1.8--2.3 solar masses, consistent with some stiff 
nuclear equations of state~\cite{vdK00}~\cite{la01}. Such masses, in comparison to the 1.4 $M_{sun}$ of some young 
neutron stars, suggest accretion has added significant mass to the neutron stars. 

\subsection{Neutron stars with high magnetic fields}
Accretion flows of neutron stars in the region where the disk
interacts with the magnetosphere have been studied through
observations of Type II x-ray bursts of the Rapid Burster~\cite{gue99} and GRO J1744$-
$28~\cite{wo00}, through low frequency QPOs in, for example, Aql X-1~\cite{rei00}, 
and through variations in the pulsed amplitude with
intensity. The latter suggest the centrifugal barrier (``propeller effect'') is operating in GRO J1744$-$28~\cite{cu97} and Aql X$-$1~\cite{cam98}~\cite{cha00}. 

\begin{figure}[thb]
\psfig{figure=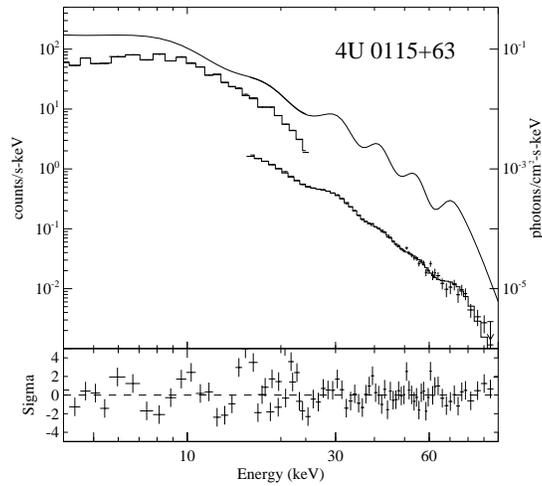,height=2.5in} 
\caption{X-ray spectrum of 4U 0115+63 showing as many as five cyclotron resonance 
absorption features, from Heindl et al. (ref. 28.).
\label{cyclo}}
\end{figure}

Cyclotron-line measurements from pulsar spectra are giving new insight into the emission geometry and magnetic fields~\cite{he99}; Fig.~\ref{cyclo}. To date, RXTE has measured the magnetic field in 10 accreting neutron star systems to be $1-5 \times 10^{12}$ G~\cite{hei00}. These values are consistent with fields inferred for older isolated neutron stars from spin-down measurements. 

The x-ray emission mechanism in rotation-powered pulsars is an unresolved issue. Phase resolved 
spectroscopy of the Crab-nebula pulsar with RXTE suggests that the emission comes 
from a small region near the magnetic poles, and not from the ``outer gap''~\cite{pra97}. The 
fastest rotation-powered pulsar J0537-6910 ($P=16$~ms) was discovered by RXTE in the 
LMC~\cite{ma98}. It is providing information about the glitches that affect young pulsars.

\begin{figure}[thb]
\psfig{figure=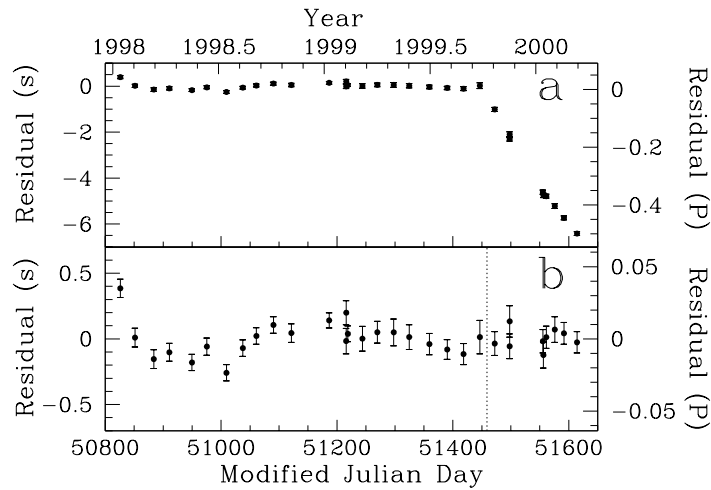,height=2.5in} 
\caption{Glitch observed in AXP 1RXS 1708--4009. The residuals from a preglitch model (a) and a post-glitch model (b) are shown, from Kaspi et al. (ref. 34).
\label{gli}}
\end{figure}

RXTE has made major contributions to the nature of anomalous x-ray pulsars (AXP) and Soft Gamma-ray Repeaters (SGR) through the study of their rapid and sometimes variable spindown rates~\cite{kou99}~\cite{mar99}~\cite{woo99}, the absence of Doppler shifts indicating the probable lack of companions~\cite{mere99}, and the earthquake-like recurrences of the outbursts in SGR~\cite{go99}. The two classes are probably closely related due to the similarity in their spin down ages $10^{3} - 10^{5}$, periods ($\sim$10 s), and luminosities ($\sim10^{35}$ erg/s)~\cite{th00}. Evidence of a glitch in one AXP, 1RXS 1708--4009~\cite{ka00}, is suggestive of kinship with other young high field pulsars; Fig.~\ref{gli}. 

Outbursts in SGR could originate in magnetic field reconnection associated with starquakes or with massive cracking of the neutron star crust in the case of the giant bursts~\cite{hur99}~\cite{fer01}. RXTE data are essential to the ongoing search for satisfactory models for the spindown and bursting of these objects~\cite{mere01}~\cite{th00}~\cite{chat00}. 

\subsection{Spectra of low-mass x-ray binaries (LMXB)}
RXTE has shown that Sco X-1 occasionally (~30\% duty cycle) exhibits non-thermal power law emission extending to beyond 200 keV at various positions along its Z track~\cite{dam01}. Less luminous (atoll) LXMBs display hard power-law emission in their low luminosity states~\cite{bar00}. These power laws indicate the presence of hot upscattering plasmas in these systems. They also complicate the spectral distinctions between neutron star and black hole systems.

\section{Transients discovered in monitoring}
RXTE continues to discover new x-ray transients with both the ASM and
PCA instruments. To the date of this conference, RXTE had found 34 previously unknown 
transients, 20 of them occurring since the beginning of 1998, some of
which are shown in Fig.~\ref{asmtr}. Several of these black-hole systems show the quasi-periodic high-frequency oscillations ($>100$ Hz) noted above (Fig.~\ref{bhosc}), that surely have origins involving general relativity.

Rapid follow-up observations of newly discovered RXTE sources with other x-ray 
observatories, and with ground based optical and radio observatories, enhance our
understanding of them. Discoveries include CI Cam 
(1998)~\cite{smCI}~\cite{be99} and V4641 Sgr (1999)~\cite{sm46}~\cite{wij00}, both of 
which were short lived and displayed associated radio emission 
indicative of relativistic jets~\cite{hjCI}~\cite{hj46}. The latter system contains a black hole 
with dynamical mass greater than 8 solar masses~\cite{oro01}. These objects are a resource 
for the continued study of black-hole systems, even in x-ray quiescence.

\section{Binary and Superorbital Periods}
The All-Sky Monitor data have revealed about 40 periodic sources with high confidence from the long-known 835-s spin period of X~Per to the previously unknown dip period of 100 days in XTE~J1716--389~\cite{we00}. Seven of these periods were previously unknown, and 4 were previously known but found for the first time in x rays with the RXTE ASM.

The long baseline of the RXTE mission is revealing superorbital 
periodicities that are relatively stable over the 5-year interval (e.g. the 60 day
period of SMC X-1) and others that evolve quite rapidly, e.g. the proposed
70 d period of Cyg X--2; both light curves are shown in Fig.~\ref{asmpr}. ASM light curves reveal an extended low-state interval in Her X-1 (Fig.~\ref{asmpr}) which could arise from a change of inclination of the precessing accretion disk~\cite{cob00}.

\section {Cataclysmic variables}
The flexibility of RXTE's rapid pointing capability has made possible timely studies of CV 
outbursts, especially short-lived rare events. The high energy response of RXTE is ideal for 
the study of high-energy shock-emitted radiation from magnetic cataclysmic variables. The 
RXTE spectra of 20 magnetic CVs yield a mean white-dwarf mass, $\sim$0.85 $M_{sun}$, 
significantly greater than do samples of isolated non magnetic white dwarfs, $\sim$0.55 $M_{sun}$~\cite{ra00}. Selection effects that could affect this result have yet to be fully 
quantified. The relation of the hard emission to the puzzling ``soft excess'' has been explored 
with coordinated RXTE/EUVE observations~\cite{chr00}. 

\section{Gamma-ray burst positions and afterglows}
RXTE plays an active role in GRB localizations and afterglow 
measurements with both the PCA and the ASM involved. At the time of this conference, the PCA had detected 5 x-ray afterglows, one of which (GRB 991216) led to an optical afterglow. Twenty 
GRB have been recorded with the ASM, with 9 of these yielding rapid positions that were 
distributed in near real time to the community. In turn, these led to four optical afterglows and 
3 redshift determinations of the 12 known at the time of this conference, including the second 
largest: $z=2.03$ for GRB~000301C. ASM x-ray light curves of GRB sometimes 
show pronounced afterpeaks not present in gamma rays that may represent the beginning of 
the afterglow~\cite{sm01}.

\section*{Acknowledgments}
The authors are grateful to all their colleagues and fellow users of RXTE, and especially to 
Drs. R. Remillard and A. Levine. We regret the omission of many pertinent references. We thank NASA for the continued operation of RXTE and for its support of RXTE science activities.



\begin{thebibliography}{99}

\bibitem{bar00} D. Barret, J. F. Olive, L. Boirin, C. Done, G. K. Skinner \& J. E. Grindlay, ApJ, 533, 329 (2000).
\bibitem{be97} T. Belloni, M. Mendez, A. R. King, M. van der Klis \& J. van der Paradijs, ApJ, 488, L109 (1997).
\bibitem{be99} T. Belloni, et al. ApJ, 527, 345 (1999).
\bibitem{blo00} P. F. Bloser, J. E. Grindlay, P. Kaaret, W. Zhang, A. P. Smale \& D. Barret, ApJ, 542, 1000 (2000).
\bibitem{br00} H. Bradt, A. Levine, R. Remillard, \& D. A. Smith, in ``X-ray Astronomy '99; Stellar Endpoints, AGN and the Diffuse Background'', Eds: G. Malaguti, G. Palumbo, and N. White, Gordon and Breach (in press), astro-ph 0003438.
\bibitem{br93} H. Bradt, R. Rothschild \& J. Swank, A\&AS, 97, 355 (1993).
\bibitem{bro99} C. Brocksopp, et al., MNRAS, 309, 1063 (1999).
\bibitem{cam98} S. Campana, L. Sella, S. Mereghetti, M. Colpi, M. Tavani, D. Ricci, D. Dal Fiume \& T. Belloni, ApJ, 499, L65 (1998).
\bibitem{ch98} D. Chakrabarty \& E. H. Morgan, Nature, 394, 346 (1998).
\bibitem{cha00} A. M. Chandler \& R. E. Rutledge, ApJ, 545, 1000 (2000).
\bibitem{chat00} P. Chatterjee \& L. Hernquist, ApJ, 543, 368 (2000).
\bibitem{chr00} D. J. Christian, AJ, 119, 1930 (2000).
\bibitem{cob00} W. Coburn, et al., ApJ, 543, 351 (2000).
\bibitem{cu97} W. Cui, ApJ, 482, L163 (1997).
\bibitem{cu98} W. Cui, S. N. Zhang \& W. Chen, ApJ, 492, L53 (1998).
\bibitem{cor00} S. Corbel, R. P. Fender, A. K. Tzioumis, M. Nowak, V. McIntyre, P. 
Durouchoux \& R. Sood, A\&A, 359, 251 (2000).
\bibitem{dam01} F. D'Amico, W. A. Heindl, R. E. Rothschild \& D. E. Gruber, ApJ, 547, L147 (2001).
\bibitem{dh00} V. Dhawan, I. F. Mirabel \& L. F. Rodriguez, ApJ, 543, 373 (2000).
\bibitem{ed99} R. Edelson \& K. Nandra, ApJ, 514, 682 (1999).
\bibitem{ei98} S. S. Eikenberry, K. Matthews, E. H. Morgan, R. A. Remillard \& R. W. 
Nelson, ApJ, 494, L61 (1998).
\bibitem{era00} M. Eracleous, R. Sambruna \& R. F. Mushotzky, ApJ, 537, 654 (2000).
\bibitem{fe99} R. Fender et al., MNRAS, 304, 865 (1999).
\bibitem{fe99a} R. Fender et al., ApJ, 519, L165 (1999).
\bibitem{fer01} M. Feroci , K. Hurley , R. C. Duncan \& C. Thompson, ApJ, 549, 1021 (2001). 
\bibitem{go99} E. Gogus, P. M. Woods, C. Kouveliotou, J. van Paradijs, M. S. Briggs, R. C. Duncan \& C. Thompson, ApJ, 526, L93 (1999). 
\bibitem{gro98} J. E. Grove, W. N. Johnson, R. A. Kroeger, K. McNaron-Brown, J. G. Skibo 
\& B. F. Phlips, ApJ, 500, 899 (1998).
\bibitem{gue99} R. Guerriero, et al., MNRAS, 307, 179 (1999).
\bibitem{he99} W. A. Heindl et al. in ``Proceedings of the 5th Compton Symposium'', 2000 (astro-ph 0004297).
\bibitem{hei00} W. A. Heindl et al., in ``Proceedings of the 5th Compton Symposium'', 2000 (astro-ph 0004300).
\bibitem{hjCI} R. M. Hjellming, IAUC 6862 (1999).
\bibitem{hj46} R. M. Hjellming, IAUC 7265 (1999).
\bibitem{hua99} X-M Hua, D. Kazanas \& W. Cui, ApJ, 512, 793 (1999).
\bibitem{hur99} K. Hurley et al., Nature, 397, 41 (1999).
\bibitem{ka00} V. M. Kaspi, J. R. Lackey \& D. Chakrabarty, ApJ, 537, L31 (2000). 
\bibitem{kaz99} D. Kazanas \& X-M Hua, ApJ, 519, 750 (1999).
\bibitem{kou99} C. Kouveliotou, et al., ApJ, 510, L115 (1999).
\bibitem{la01} F. K. Lamb \& M. C. Miller, ApJL, submitted (astro-ph 0007460).
\bibitem{mar99} D. Marsden, R. E. Rothschild, \& R. E. Lingenfelter, ApJ, 520, L107 (1999). 
\bibitem{ma98} F. E. Marshall, E. V. Gotthelf, W. Zhang, J. Middleditch \& Q. D. Wang, 
ApJ, 499, L179 (1998).
\bibitem{mere01} S. Mereghetti in``Frontier Objects in Astrophysics and Particle Physics'', Vulcano, Italy, 2000 (astro-ph 0102017). 
\bibitem{mere99} S. Mereghetti, G. L. Israel \& L. Stella, MNRAS, 296, 689 (1999).
\bibitem{mer99} A. Merloni, M. Vietri, L. Stella \& D. Bini, MNRAS, 304, 155 (1999).
\bibitem{mi98} M. C. Miller, F. K. Lamb \& G. B. Cook, ApJ, 508, 791 (1998).
\bibitem{mir94} I. F. Mirabel \& L. F. Rodriguez, Nature, 371, 46 (1994).
\bibitem{mir98} I. F. Mirabel, et al., A\&A, 330, L9 (1998).
\bibitem{mir99} F. Mirabel \& L. Rodriguez, ARAA, 37, 409 (1999).
\bibitem{mun99} M. Muno, E. H. Morgan \& R. A. Remillard, ApJ, 527, 321 (1999).
\bibitem{na95} R. Narayan \& I. Yi, ApJ, 452, 710 (1995).
\bibitem{na01} K. Nandra, Adv. in Space Res., in press (astro-ph 0012448).
\bibitem{no97} M. Nowak, R. Wagoner, R. Begelman, M. Lehr \& E. Dana, ApJ, 477, L91 (1997).
\bibitem{now99} M. A. Nowak, B. A. Vaughan, J. Wilms, J. B. Dove \& M. C. Begelman, 
ApJ, 510, 874 (1999).
\bibitem{oro01} J. A. Orosz, et al., ApJ, submitted (astro-ph 0103045).
\bibitem{poo97} G. G. Pooley \& R. P. Fender, MNRAS, 292, 925 (1997).
\bibitem{po99} J. Poutanen \& A. C. Fabian , MNRAS, 306, L31 (1999).
\bibitem{pra97} S. H. Pravdo, L. Angelini \& A. K. Harding, ApJ, 491, 808 (1997).
\bibitem{ra00} G. Ramsay, MNRAS, 314, 403 (2000).
\bibitem{rei00} P. Reig, M. Mendez, M. van der Klis \& E. C. Ford, ApJ, 530, 916 (2000).
\bibitem{rem300} R. Remillard, E. Morgan, C. Bailyn \& J. Orosz, ApJ, 522, 397 (1999).
\bibitem{remBH} R. Remillard, in ``Evolution of Binary and Multiple Stars'', Bormio, Italy, June 2000, eds. P. Podsiadlowski, S. Rappaport, A. King, F. D'Antona \& L. Burderi (San Francisco: ASP), in press (astro-ph 0103431).
\bibitem{rot99} R. E. Rothschild et al., ApJ, 510, 651 (1999).
\bibitem{rey00} C. S. Reynolds, ApJ, 533, 811 (2000).
\bibitem{sa00} R. M. Sambruna, et al., ApJ, 538, 127 (2000).
\bibitem{smCI} D. A. Smith, et al., IAUC 6855 (1999).
\bibitem{sm46} D. A. Smith, et al., IAUC 7253 (1999).
\bibitem{sm01} D. A. Smith, et al., ApJ, submitted (astro-ph. 0103357).
\bibitem{st99} L. Stella, M. Vietri \& S. Morsink, ApJ, 524, L63 (1999).
\bibitem{sti97} A. Stirling, R. Spencer \& M. Garrett, New Astron. Rev., 42, 657 (1998).
\bibitem{str96} T. Strohmayer, W. Zhang, J. H. Swank, A. Smale, L. Titarchuk, C. Day \& U. 
Lee, ApJ, 469, L9 (1996).
\bibitem{th00} C. Thompson, in ``Proceedings NATO Advanced Study Institute: The Neutron Star-Black Hole Connection'', Elounda, Crete, 1999, eds. V. Connaughton, C. Kouveliotou, J. van Paradijs, J. Ventura (astro-ph 0010016).
\bibitem{vdK00} M. van der Klis, ARAA, 38, 717 (2000).
\bibitem{vaP00} J. van Paradijs, C. Kouveliotou \& R. Wijers, ARAA, 38, 379 (2000). 
\bibitem{we00} L. Wen, PhD thesis, MIT, (2000).
\bibitem{wi98} R. Wijnands \& M. van der Klis, Nature, 394, 344 (1998).
\bibitem{wij00} R. Wijnands \& M. van der Klis, ApJ, 528, L93 (2000).
\bibitem{woo99} P. M. Woods et al., ApJ, 524, L55 (1999).
\bibitem{wo00} P. M. Woods et al., ApJ, 540, 1062 (2000).
\bibitem{zh97} S. N. Zhang, W. Cui, W. Chen, ApJ, 482, L155 (1997).
\bibitem{zh98} W. Zhang, A. P, Smale, T. E. Strohmayer \& J. H. Swank, ApJ, et al., ApJ, 500, L171 (1998).
\end{thebibliography}
\end{document}